\documentclass[aps,prx,reprint,longbibliography,superscriptaddress,floatfix]{revtex4-2}
\usepackage{symbols}
\usepackage{xfrac}
\usepackage{amsfonts}
\usepackage{dsfont}
\usepackage{amsmath}
\usepackage{amssymb}
\usepackage{amsthm}
\usepackage{mathtools}
\usepackage{graphicx}
\usepackage[caption=false]{subfig}
\usepackage{floatrow}
\usepackage{color}
\usepackage{bm}
\usepackage[normalem]{ulem}
\usepackage[hidelinks]{hyperref}
\hypersetup{
     colorlinks   = true,
     citecolor    = blue,
     linkcolor    = blue
}
\usepackage{mathptmx}
\usepackage{enumitem}

\usepackage{relsize}
\usepackage{xcolor}
\usepackage{scalerel}
\usepackage{tikz}
\usetikzlibrary{svg.path}
\usepackage{soul}
\definecolor{orcidlogocol}{HTML}{A6CE39}
\tikzset{
  orcidlogo/.pic={
    \fill[orcidlogocol] svg{M256,128c0,70.7-57.3,128-128,128C57.3,256,0,198.7,0,128C0,57.3,57.3,0,128,0C198.7,0,256,57.3,256,128z};
    \fill[white] svg{M86.3,186.2H70.9V79.1h15.4v48.4V186.2z}
                 svg{M108.9,79.1h41.6c39.6,0,57,28.3,57,53.6c0,27.5-21.5,53.6-56.8,53.6h-41.8V79.1z M124.3,172.4h24.5c34.9,0,42.9-26.5,42.9-39.7c0-21.5-13.7-39.7-43.7-39.7h-23.7V172.4z}
                 svg{M88.7,56.8c0,5.5-4.5,10.1-10.1,10.1c-5.6,0-10.1-4.6-10.1-10.1c0-5.6,4.5-10.1,10.1-10.1C84.2,46.7,88.7,51.3,88.7,56.8z};
  }
}
\newcommand\orcid[1]{\href{https://orcid.org/#1}{\mbox{\scalerel*{
\begin{tikzpicture}[yscale=-1,transform shape]
\pic{orcidlogo};
\end{tikzpicture}
\ 
}{|}}}}

\newcommand{\red}[1]{{\color{red} {#1}}}
\newcommand{\stkout}[1]{\ifmmode\red{\text{\sout{\ensuremath{#1}}}}\else \red{\sout{#1}}\fi}
\begin{document}

\title{Certifying Macroscopic Quantum Mechanics via Hypothesis Testing with Finite Data}

\author{Andreu Riera-Campeny \orcid{0000-0003-3260-993X}}
\affiliation{ICFO – Institut de Ciencies Fotoniques, The Barcelona Institute of Science and Technology, Castelldefels, Barcelona 08860, Spain}
\affiliation{Institute for Quantum Optics and Quantum Information of the Austrian Academy of Sciences, 6020 Innsbruck, Austria}

\author{Patrick Maurer \orcid{0000-0002-0423-8898}}
\affiliation{Institute for Quantum Optics and Quantum Information of the Austrian Academy of Sciences, 6020 Innsbruck, Austria}

\author{Oriol Romero-Isart \orcid{0000-0003-4006-3391}}
\affiliation{ICFO – Institut de Ciencies Fotoniques, The Barcelona Institute of Science and Technology, Castelldefels, Barcelona 08860, Spain}
\affiliation{ICREA – Institucio Catalana de Recerca i Estudis Avançats, Barcelona 08010, Spain}

\date{\today}

\begin{abstract}
We address the challenge of certifying quantum behavior with single macroscopic massive particles, subject to decoherence and finite data.
We propose a hypothesis testing framework that distinguishes between classical and quantum mechanics based on position measurements.
While interference pattern visibility in single-particle quantum superposition experiments has been commonly used as a sufficient criterion to falsify classical mechanics, we show that, from a hypothesis testing perspective, it is neither necessary nor efficient.
Focusing on recent proposals to prepare macroscopic superposition states of levitated nanoparticles, we show that the likelihood ratio test---which leverages differences across the entire probability distribution--- provides an exponential reduction in measurements needed to reach a given confidence level.
These results generalize to a broad class of quantum states, and offer a principled, efficient method to falsify classical mechanics in interference experiments, relaxing the experimental constraints faced by current efforts to test quantum mechanics at the macroscopic scale.
\end{abstract}

\maketitle

\section{Introduction}
The gradual emergence of an interference pattern from successive position measurements of single massive particles in a double-slit experiment remains one of the most striking demonstrations of the quantum superposition principle~\cite{Tonomura1989, Juffmann2012}.
The appearance of such an interference pattern in the position probability distribution, under conditions where classical mechanics cannot account for it, provides a sufficient criterion to certify quantum mechanics~\cite{Arndt1999, Brezger2002, Hornberger2012, Eibenberger2013, Kovachy2015, Fein2019}.
Pushing quantum interference experiments to increasingly macroscopic scales, both in mass and in delocalization distance, offers a way to probe the boundaries of quantum mechanics and test potential modifications of the theory~\cite{RomeroIsart2011, Nimmrichter2013, Bassi2013, Arndt2014}.
However, this endeavor comes at a cost. Larger physical systems are more susceptible to the interaction with their environment, making the visibility of their delicate quantum interference more difficult to observe~\cite{Joos1985, Zurek2007, Schlosshauer2007}. 
Reliably certifying quantum mechanics with macroscopic particles therefore requires mitigating environmental decoherence, in-run noise, and run-to-run experimental fluctuations~\cite{Hornberger2012}.
While recent experimental efforts with levitated particles have focused on reducing noise and decoherence~\cite{Dania2023, Bonvin2024, Rossi2025, Lindner2024, Dago2024, Tomassi2025}, the certification process itself has received less attention.

In this work, we focus precisely on this second aspect: the certification process in experiments aiming to prepare macroscopic quantum superposition states of a single massive particle. 
More specifically, we frame the problem as a binary hypothesis test~\cite{Tsang2013} where, given a set of $\nm$ position measurements, we aim to distinguish between two hypotheses: the null hypothesis $\hypothesis_0$, consistent with classical mechanics, and the alternative hypothesis $\hypothesis_1$, which assumes quantum mechanical behavior.
Framing the certification problem in this way allows us to compare different statistical tests---such as the likelihood ratio test and the visibility test---and to analyze their performance as a function of the number of measurements $\nm$ and the level of decoherence.

We apply this framework to recent experimental proposals involving levitated nanoparticles~\cite{RodaLlordes2023, Neumeier2024}, a promising platform for testing the superposition principle at unprecedented mass and delocalization scales. Furthermore, we show that our findings extend to a wide range of well-defined macroscopic quantum states. The center-of-mass motion of these systems has been experimentally cooled to the quantum ground state~\cite{Delic2020, Magrini2021, Tebbenjohanns2021, Ranfagni2022, Kamba2022, Piotrowski2023}, and mildly delocalized beyond its zero-point fluctuations~\cite{Rossi2025, Kamba2025}. 
For the dynamical protocols in these proposals, we show that, remarkably, the likelihood ratio test---the optimal hypothesis testing method---could exponentially reduce the number of measurements needed to certify quantum mechanics at a given significance level, outperforming conventional interference visibility tests. As we shall argue, these results can open the door to a decoherence-tolerant certification process which requires fewer measurements, thereby reducing also the impact of run-to-run variability.

Our article is organized as follows. In Section~\ref{sec:model} we define the theoretical model for the classical and quantum dynamics of single macroscopic massive particles in the presence of noise and decoherence. In Sec.~\ref{sec:framework}, we introduce the hypothesis-testing framework to certify quantum behavior with finite position measurement data. In Sec.~\ref{sec:application}, we apply this framework to the specific case of cubic phase state---a standard resource in protocols for creating macroscopic quantum superpositions with levitated nanoparticles---and demonstrate the exponential advantage of the likelihood ratio test in Sec.~\ref{sec:results}. In Sec.~\ref{sec:interpretation}, we provide a physical interpretation of this advantage and formally demonstrate that the contrasting scaling laws are general for a broad class of semiclassical states in Sec.~\ref{sec:universality}. Finally, in Sec.~\ref{sec:conclusion} we provide concluding remarks.

\section{Model and Open-System Dynamics}
\label{sec:model}
We consider the one-dimensional center-of-mass motion of a massive particle with mass $\mass$, described by position $\posop$ and momentum $\momop$ operators, which satisfy the canonical commutation relation $[\posop, \momop] = i \hbar$. At time $t = 0$, the particle is confined in a harmonic potential (e.g., an optical trap) with frequency $\trapfreq$ and is prepared in a thermal state with a mean phonon occupation number $\nbar$. We denote by $\zppos = \sqrt{\hbar/(2\mass\trapfreq)}$ and $\zpmom=\sqrt{\hbar\mass\trapfreq/2}$ the zero point fluctuations in this potential. 
For $t>0$, the system is subjected to a dynamical protocol implemented via a possibly time-dependent potential $\pot(\pos,t)$ and evolves according to either classical or quantum equations of motion in the presence of noise and decoherence.
For massive particles in highly isolated environments without damping, the dominant sources of decoherence are well captured by a momentum diffusion process~\cite{RomeroIsart2011}, whose diffusion coefficient we denote by $\diffrate$. 
The distinction between classical and quantum dynamics is highlighted in the evolution equation for the Wigner function, which takes the form
\begin{multline}\label{eq:wigner_equation}
    \frac{\partial \wigner_s}{\partial t}(\pos,\mom,t) = \left\{ - \frac{\mom}{\mass} \frac{\partial}{\partial \pos} + \frac{\partial \pot}{\partial \pos}(\pos,t) \frac{\partial}{\partial\mom} + \frac{\diffrate}{2} \frac{\partial^2}{\partial\mom^2} \right. \\
    \left. + s \sum_{n\geq1} \frac{\left(i\hbar/2\right)^{2n}}{(2n+1)!}\frac{\partial^{2n+1} \pot}{\partial \pos^{2n+1}}(\pos,t) \frac{\partial^{2n+1}}{\partial\mom^{2n+1}} \right\} \wigner_s(\pos,\mom,t),
\end{multline}
where $s\in \{0,1\}$ labels the classical or quantum hypothesis, respectively. In Eq.~\eqref{eq:wigner_equation}, the first line corresponds to the classical dynamics, including kinetic and potential terms that describe the Hamiltonian flow, along with the diffusive term accounting for noise and decoherence.
The second line, proportional to $s$, corresponds to the quantum term. Note that, in the absence of decoherence, for $\diffrate = 0$, Eq.~\eqref{eq:wigner_equation} reduces to the Liouville equation for $s = 0$ and to the Schr\"odinger equation for $s = 1$.
The initial condition $\wigner(\pos,\mom,0) = \wigner_\mathrm{i}(\pos,\mom)$ is independent of the hypothesis, and corresponds to the aforementioned thermal state, with variances $\variance_\pos = (2\nbar+1)\zppos^2$ and $\variance_\mom = (2\nbar +1)\zpmom^2$. 
After an evolution time $t$, the classical $\pdfcl(\pos)$ or quantum $\pdfqt(\pos)$ position probability distribution can be calculated as $\pdf_s(\pos) = \int_{-\infty}^\infty \wigner_s(\pos,\mom,t) d\mom$. These probability distributions will depend on a set of parameters, labeled as $\params = (\param_1, \cdots, \param_\np)$, which are determined, for instance, from the properties of the potential $\pot(\pos,t)$ defining the protocol.

Having defined the theoretical model for the classical and quantum dynamics, we now introduce the statistical framework used to distinguish between the two hypotheses using finite data.

\section{Certification via Hypothesis Testing}
\label{sec:framework}
After the protocol terminates at time $t$, a measurement is performed on the system, yielding an outcome $\datapoint_1$. We model the measurement using a standard linear model, $\datapoint_1 = \signal_1 + \noise_1$, where $\signal_1$ is the true measurement outcome (e.g., the particle position) and $\noise_1$ is a zero-mean Gaussian random variable with variance $\variance_R$ that represents random measurement noise. This description naturally captures the retrodiction of measurement outcomes in continuously monitored systems~\cite{Zhang2017, Lammers2024}. The state of the massive particle is then re-initialized, and the protocol is repeated $\nm$ times, generating a dataset $\{\datapoint_1, \cdots, \datapoint_\nm\}$. The measurement outcomes are assumed to be independent and sampled from the distribution $\pdf_s(\datapoint)$, with $s\in \{0,1\}$ corresponding to either classical or quantum dynamics. 

\subsection{Test Statistics: Visibility and Likelihood-Ratio}

\begin{figure}[h]
    \includegraphics[width=\linewidth]{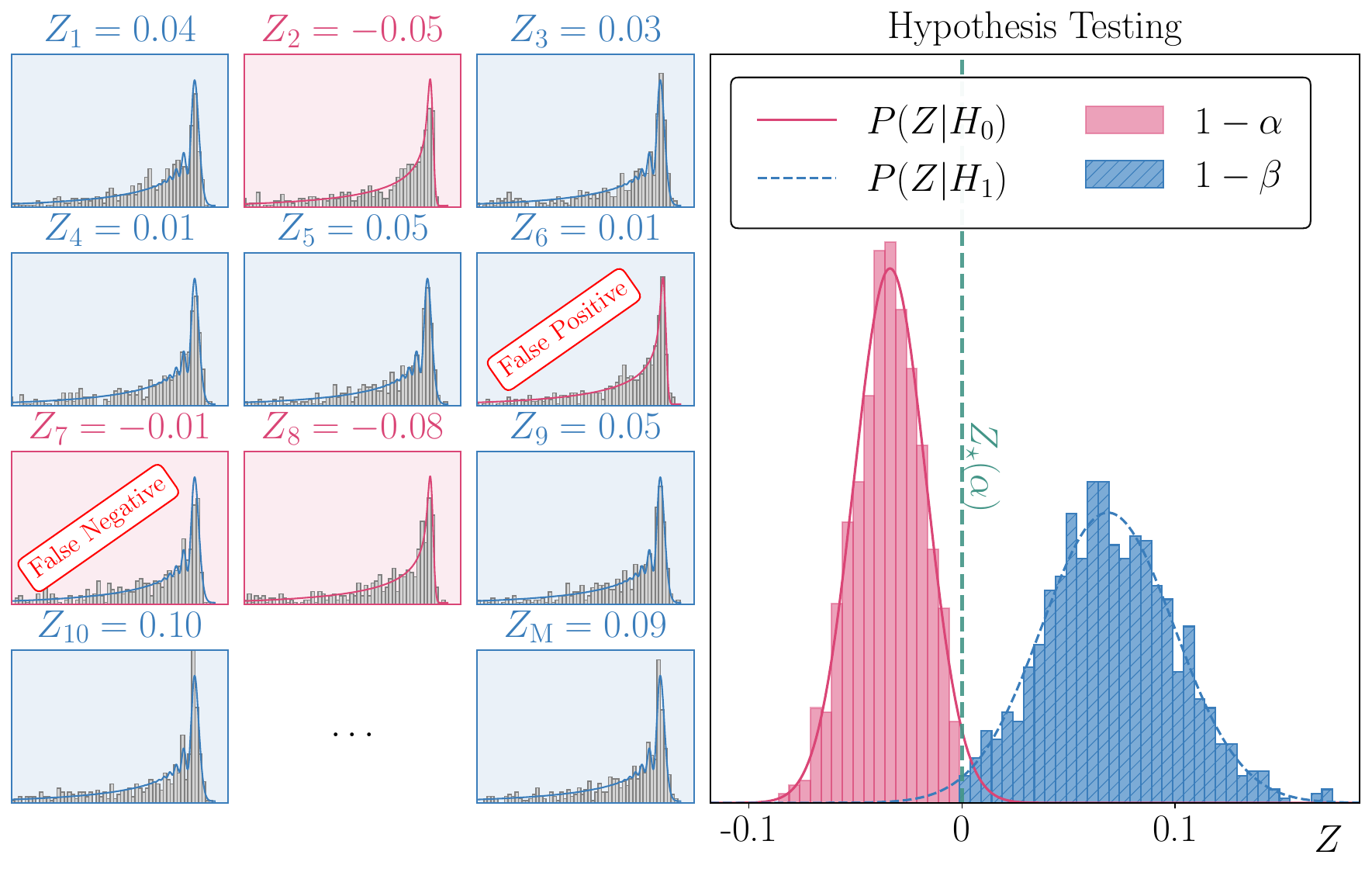}
\caption{(Left) Schematic representation of $M=5000$ classical (red solid) and quantum (blue solid) datasets $\{\datapoint_1, \cdots, \datapoint_\nm\}$ with $\nm = 500$, from which histograms (gray) are constructed. A test statistic $Z_i$ is computed for each instance and compared against a decision threshold $Z_\star(\sig) = 0$ (for notational convenience). Instances with $Z_i <0$ (red background) are classified as classical; that is, $\hypothesis_0$ is accepted, while those with $Z_i > 0$ (blue background) are classified as quantum, corresponding to acceptance of $\hypothesis_1$. (Right) Empirical distribution of $Z$ values superimposed with theoretical distributions $\prob(Z|\hypothesis_0)$ and $\prob(Z|\hypothesis_1)$ that correspond to the prediction of classical (red, solid) and quantum (blue, hatched) mechanics, respectively. The area to the left of the threshold and under the $\prob(Z|\hypothesis_0)$ curve corresponds to the confidence level $1-\sig$. The area to the right of the threshold and under the $\prob(Z|\hypothesis_1)$ curve corresponds to the test power $1-\beta$. 
}
    \label{fig:schematics}
\end{figure}

To test our hypothesis, we use a random variable $Z$, the so-called test statistic, which is a function of the measurement data $Z = f(\datapoint_1, \cdots, \datapoint_\nm)$. In the limit of infinite data, $\nm \to \infty$, any difference between the predictions of quantum and classical mechanics can be used to certify quantum behavior with certainty. However, with finite data, a threshold must be selected to decide when a null hypothesis is rejected. The threshold is typically determined by fixing the significance level $\sig$, which is defined as the probability of rejecting the null hypothesis $\hypothesis_0$ when the dataset $\{\datapoint_1, \cdots, \datapoint_\nm\}$ is actually generated under the null hypothesis; that is, a false positive incorrectly suggesting that classical mechanics has been falsified. Once the significance level is set, the statistical power of the test can be evaluated. It is defined as 
$1-\beta$, and corresponds to the probability of correctly rejecting the null hypothesis~\cite{Wasserman2004}, that is, correctly falsifying classical mechanics. See Fig.~\ref{fig:schematics} for further details. Hence, the test power can be used as a figure of merit to evaluate the performance of a statistical test given a desired significance level $\sig$. Let us now discuss two test statistics: the visibility test and the likelihood ratio test.

The visibility test, defined by the random variable $Z = \vis$, is given by the ratio
\begin{align}
\vis =
\frac{\nmax - \nmin}{\nmax + \nmin}, \label{eq:visibility}
\end{align}
where $\nmax$ and $\nmin$ are the counts in the bins corresponding to the maximum $\posmax$ and minimum $\posmin$ of the interference pattern associated with the quantum probability distribution. This statistic provides a measure of the local contrast near the minimum of the interference pattern. The specific procedures used to determine these values, along with further details about this definition, are detailed in Appendix~\ref{app:visibility_test}. With infinite data, the number of counts converges to the true probabilities. Therefore, in the presence of an interference pattern for the quantum distribution---absent under the classical hypothesis--- one could certify quantum mechanics with the visibility test by observing $\vis>0$.

The (log-) likelihood ratio test $Z=\lrt$ of a set of independent measurements is defined as the logarithm of the ratio of the likelihood of the data under the alternative hypothesis to the likelihood of the data under the null hypothesis, namely,
\begin{align}
    \lrt = \frac{1}{\nm} \sum_{i=1}^\nm \log\left[\frac{\pdfqt(\datapoint_i)}{\pdfcl(\datapoint_i)}\right],~\label{eq:likelihoodratiotest}
\end{align}
where the factor $1/\nm$ is added for normalization. Neyman-Pearson's lemma ensures that the likelihood ratio test $\lrt$ is optimal for hypothesis testing under mild assumptions~\cite{Cover2005}. Since the likelihood ratio test is a sum of independent and identically distributed random variables, the central limit theorem guarantees that it converges to a Gaussian random variable of decaying variance as the number of measurements $\nm$ goes to infinity. The expectation value of the distribution is given by 
\begin{equation}
     \av{\lrt}_s = \begin{cases}
        \relentropy{\pdfqt}{\pdfcl} & \mathrm{for} \quad s=1 \\
        -\relentropy{\pdfcl}{\pdfqt} & \mathrm{for} \quad s=0
    \end{cases},
\end{equation}
where $\av{\cdot}_s$ is the average according to $\pdf_s(Y)$, and $\relentropy{\pdf}{\qdf} = \int \pdf(x) [\log(\pdf(x))-\log (\qdf(x))] dx \geq 0$ is the non-negative relative entropy, with equality only if $\pdf = \qdf$. Therefore, given a sufficiently large number of measurements $\nm$ and that $\pdfqt(\datapoint) \neq \pdfcl(\datapoint)$, quantum mechanics could be always certified using the likelihood ratio test by observing $\lrt > 0$. We note in passing that hypothesis testing across a continuum of hypotheses has been used  to quantify the degree of macroscopicity of quantum superpositions~\cite{Nimmrichter2013, Schrinski2019, Schrinski2020}. This approach, however, should be distinguished from the binary certification problem considered here.

\subsection{Performance Metric} 
To compare the two statistical tests, we compute the number of measurements $\nmt$ required to falsify classical mechanics at a significance level $\sig = 2.86 \times 10^{-7}$---corresponding to the conventional 5-sigma threshold---and with statistical power $1-\beta \geq 0.9973$. In other words, we determine the minimal $\nmt$ such that the probability of falsely identifying classical data as quantum is less than one in a million, while correctly identifying quantum data exceeds $99.73\%$. This metric provides a direct and quantitative measure of a test’s efficiency in distinguishing classical from quantum dynamics based on the position distribution at a specified level of statistical confidence. 

\section{Case Study: Cubic Phase States}
\label{sec:application}
To apply and analyze the proposed certification method, we focus on dynamical protocols $\pot(\pos,t)$ for creating macroscopic quantum superpositions with a single, initially cooled, levitated nanoparticle in a wide non-harmonic potential~\cite{Neumeier2024, RodaLlordes2023}. These protocols commonly lead to the preparation of cubic phase states~\cite{RieraCampeny2024}. While we use them as a primary example, we show in Sec.~\ref{sec:universality} that the following analysis extends to the entire class of states for which quantum mechanics is imprinted as a high-frequency feature of the wavefunction under a slowly varying envelope. These states arise naturally in the Wentzel–Kramers–Brillouin (WKB) approximation and in related semiclassical constructions \cite{Berry1972, LandauLifshitz1977, Heller2018}. 

\subsection{Key Properties}

Cubic phase states~\cite{Gottesman2001, Kala2022, Moore2025} are characterized by an oscillatory behavior on one side of the position distribution's maximum and a rapid decay on the opposite side, see inset of Fig.~\ref{fig:robustness}. As shown in Appendix~\ref{app:cubic_states}, the associated characteristic function $\charfunc_s(k) = \int \pdf_s(\datapoint) \exp(i k \datapoint) d\datapoint$, in one-to-one correspondence with the probability distribution, is given by
\begin{align}
    \charfunc_s(k) = \dfrac{\exp(-i s \param_3 k^3/3 - \param_2 k^2/2)}{ \sqrt{1+2i \param_1 k} }, \label{eq:charfunc_cubic_s}
\end{align}
for $s \in \{0,1\}$. The parameters $\param_1$, $\param_2$, and $\param_3$ are real numbers subject to $\param_2>0$ and $\param_3/(\param_2 \param_1) \in [0,1]$, which ensures the positivity of their associated quantum state. 
The role of the parameters $\param_1$, $\param_2$, and $\param_3$ can be understood from their appearance in the expression of the cumulants $\cumulant_{s,j}$ of the distribution $\pdf_s(\datapoint)$ associated to $\charfunc_s(k)$. We find that $\mu_s = \cumulant_{s,1} = -\param_1$ where $\mu_s$ is the average of the distribution, and $\variance_s = \cumulant_{s,2} = \param_2 + 2\param_1^2$, where $\variance_s$ is the variance. Finally, the parameter $\param_3$ is associated to the skewness (i.e., the third cumulant) of the distribution as $\cumulant_{s,3} = 2s\param_3 - 8\param_1^3$. Interestingly, the third cumulant is the only one that contains information about the hypothesis $s$, and the quantum contribution has the effect of reducing the classical skewness; that is, it makes the distribution more symmetric. Higher order cumulants depend only on the parameter $\param_1$ as $\cumulant_{s,j} = (j-1)! (-2\param_1)^j/2$, with $j \geq 4$. Finally, we note that the purity of the associated quantum state can be expressed as $\sqrt{\param_3/(\param_2 \param_1)}$.

Cubic phase states display different quantum and classical measurement statistics. This distinction is highlighted by the fact that their associated distributions are related through the Airy transform~\cite{vallee_airy_2010}
\begin{align}
\pdfqt(\datapoint) = \frac{1}{|\sqrt[3]{\param_3}|} \int_{-\infty}^\infty \Airy\left(\frac{\datapoint - \datapoint'}{\sqrt[3]{\param_3}} \right) \pdfcl(\datapoint') d\datapoint', \label{eq:airy_transform}
\end{align}
where $\Airy(x)$ denotes the Airy function. Hence, the quantum and classical distributions remain distinct for any finite $\param_3$, coinciding only in the limit $\param_3 \to 0$. 
We are interested in how this difference depends on noise and decoherence. To quantify their impact on position measurements, we express the observed distributions $\pdf_s(\datapoint)$---both classical and quantum---as convolutions of the ideal noiseless distributions $\tilde{\pdf}_s(\signal)$, corresponding to $\nbar = 0$, $\diffrate = 0$, and $\variance_R = 0$, with a Gaussian kernel. This yields
\begin{equation}
\label{eq:low_pass}
\pdf_s(\datapoint) = \frac{1}{\sqrt{2\pi\sigma^2}} \int_{-\infty}^\infty \exp\left[-\frac{(\datapoint - \signal)^2}{2\sigma^2}\right] \tilde{\pdf}_s(\signal) d\signal,
\end{equation}
where $\sigma^2 = (\param_2 - \param_3/\param_1) + \variance_R$ quantifies the effect of noise and decoherence on the measured data. 

Having established the theoretical description of cubic phase states established, we proceed to a performance analysis to quantify the efficiency of the certification frameworks.

\subsection{ Scaling Analysis of $N_\star$}
\label{sec:results}
\begin{figure}[t]
    \centering
    \includegraphics[width=1.\linewidth]{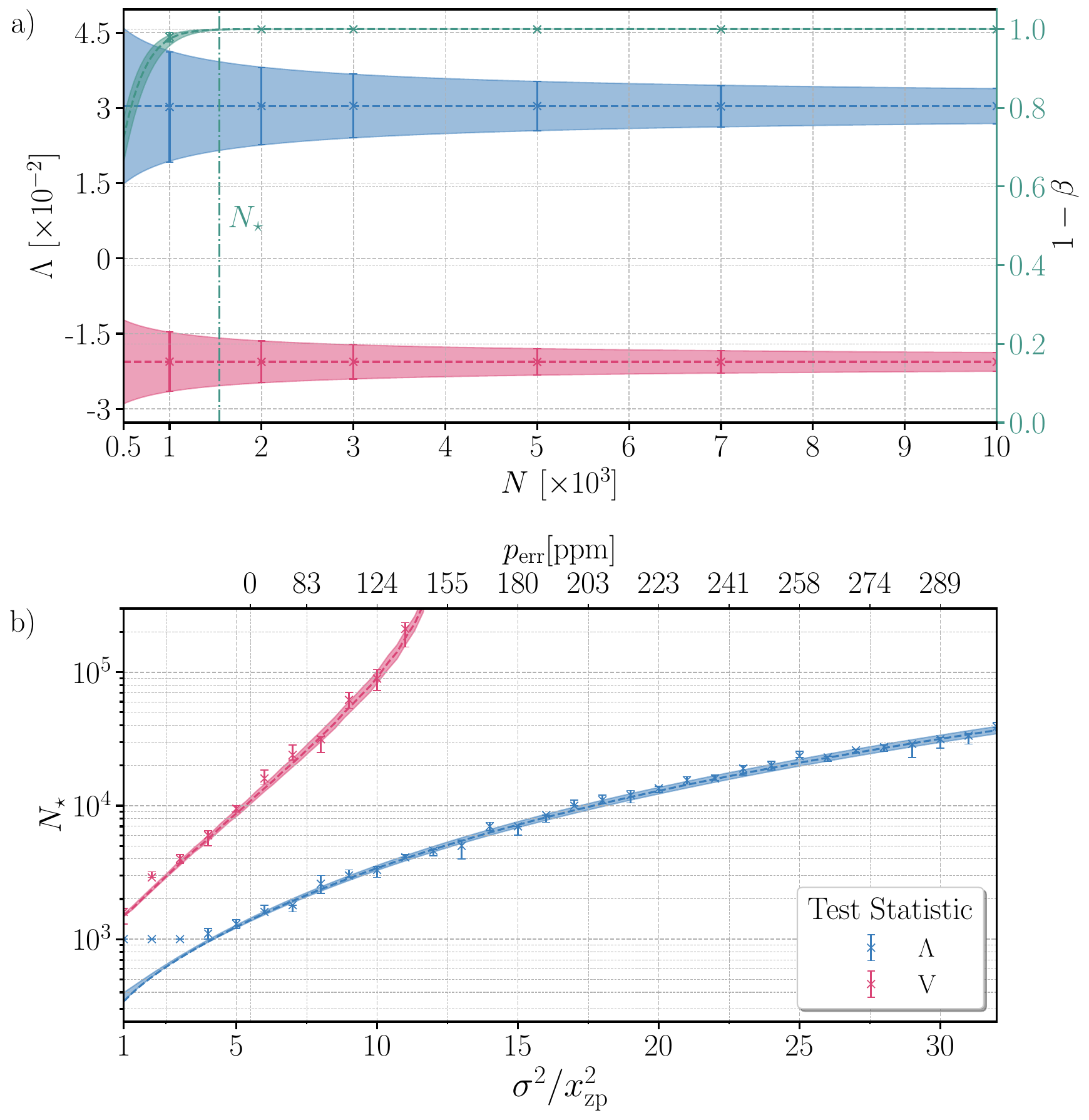}
     \caption{
     (a) Mean (symbols) and standard deviation (error bars) of the distribution of $\lrt$ simulated over $M=5000$ runs each for the classical (red) and quantum (blue) hypothesis as a function of the number of measurements $N$ using the parameter values specified in Tab.~\ref{tab:sm_params} of Appendix~\ref{app:physical_parameters}. Dashed lines and shaded regions represent the mean value and standard deviation obtained by integration in the large $N$ limit. Green symbols show the (empirical) statistical power $1-\beta$ for a significance level $\sig = 2.86\times 10^{-7}$, whereas the dashed green line shows the statistical power obtained by integration in the asymptotic limit. Green error bars and shaded regions correspond to a $\pm 5\%$ uncertainty in $\sigma^2$ and $\param_3$. The vertical dash-dotted line shows $\nmt$, i.e. the number of measurements $N$ for which the power is above $99.73\%$ for all parameters in the $\pm5\%$ error window.  (b) Number of measurements $\nmt$ needed to certify quantum mechanics as a function of $\sigma^2$ (lower $x$-axis) and the corresponding relative precision error $p_\text{err}$ in the pulse energy (upper $x$-axis) as defined in Ref.~\cite{Neumeier2024}. The symbols and error bars correspond to a statistical power $1-\beta>99.73\%$ at a significance level $\sig = 2.86\times 10^{-7}$ for the visibility (red) and likelihood-ratio (blue) test statistics obtained by $M=2000$ runs with $\param_3$ within a $\pm5\%$ error window of the value specified in Tab.~\ref{tab:sm_params} of Appendix~\ref{app:physical_parameters}. The corresponding dashed lines and shaded regions show the values obtained by integration in the large $N$ limit. Details about numerical methods can be found in Appendix~\ref{app:details_numerics}.}
    \label{fig:2}
\end{figure}

Focusing on the protocols that generate the cubic phase states discussed above, we now discuss the results of our certification method. In Fig.~\ref{fig:2}a, we show the convergence of the likelihood ratio test statistic $\lrt$ toward the relative entropy as a function of the number of measurements $\nm$, for cubic phase states and using the parameter values specified in Tab.~\ref{tab:sm_params} of the Appendix~\ref{app:physical_parameters}, corresponding to Ref.~\cite{Neumeier2024}. The rapid convergence with $\nm$ ensures that the likelihood ratio test can be used to discriminate between the classical and quantum scenarios with a $5-$sigma significance level already for $\nm_\star \approx 1500$ measurements. We note that such a $\nmt$ will always exist for the likelihood ratio test, since the variance of the likelihood ratio test decays as $1/\nm$ for sufficiently large $\nm$. Indeed, according to Eq.~\eqref{eq:airy_transform}, the two probability distributions $\pdfcl(\datapoint)$ and $\pdfqt(\datapoint)$ are different and thus always distinguishable with a sufficiently large dataset.

While the likelihood ratio test can always distinguish between the classical and quantum hypotheses given sufficient data, this task becomes increasingly challenging as the decoherence increases. In Fig.~\ref{fig:2}b, we compare the metric $\nmt$ for both the likelihood ratio test and the visibility test as a function of $\sigma^2$. As shown, the likelihood ratio test consistently outperforms the visibility test across all plotted decoherence values. The difference between the two tests is highlighted by the scaling behavior of $\nmt$ with $\sigma^2$. While the likelihood ratio test has a polynomial scaling with decoherence, the visibility test exhibits exponential growth. Thus, using the likelihood ratio test as a test statistic leads to an exponentially more efficient testing.
In turn, this improved efficiency enables certification under less stringent experimental conditions. For example, for the proposal of Ref.~\cite{Neumeier2024}, the visibility test requires pulse energy precision $p_\text{err} \lesssim 10^{-5}$ for $\nmt \approx 10^4$, whereas the likelihood ratio test relaxes this limiting constraint by over an order of magnitude. This raises the question of the physical origin of these results and whether they are an artifact of the specific state under consideration. We address this in Sec.~\ref{sec:discussion}, where we identify the mechanism as a low-pass filtering effect of decoherence and prove its universality for a broad class of well-defined states.

\section{Discussion}
\label{sec:discussion}

\subsection{Decoherence as a Low-Pass Filter}
\label{sec:interpretation}

\begin{figure}[h]
    \centering
    \includegraphics[width=1.\linewidth]{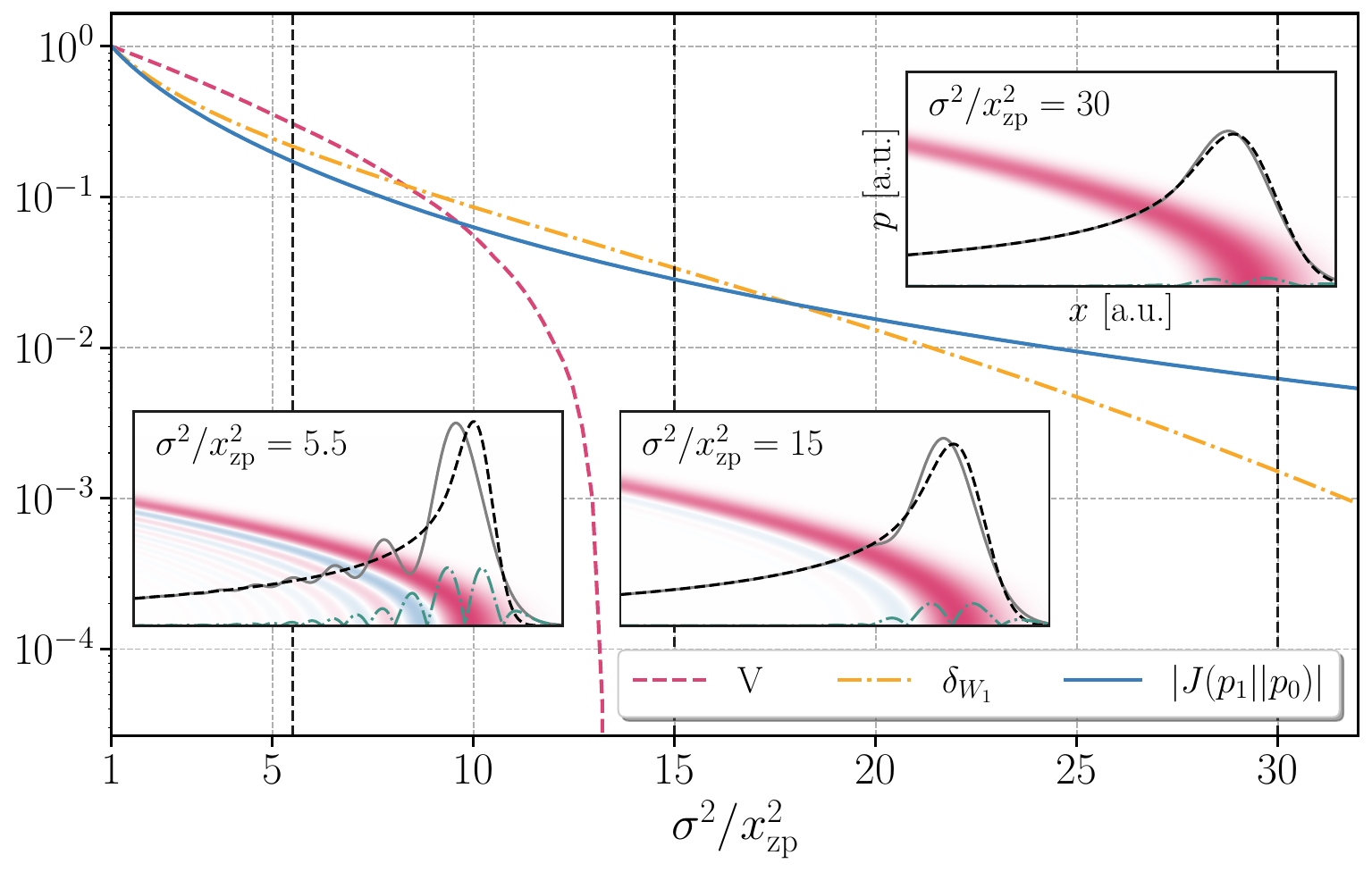}
     \caption{Visibility $\vis$ (red, dashed), Wigner function negativity $\delta_{W_1} = \int |W_1(x,p)|dxdp-1$ (orange, dash-dotted) and Jeffreys divergence  $\jeffreysdiv{\pdfqt}{\pdfcl}$  (blue, solid) as a function of $\sigma^2$ for $\param_3$ as specified in Tab.~\ref{tab:sm_params} of Appendix~\ref{app:physical_parameters}. All functions are normalized to their value at $\sigma^2=\zppos^2$. The insets show the quantum and classical probability distribution $p_1(x)$ (gray, solid), $p_0(x)$ (black, dashed), their absolute difference $|p_1(x)-p_0(x)|$ (green, dash-dotted) and the Wigner function $W_1(x,p)$ (red and blue shaded regions represent positive and negative values of the Wigner function, respectively) as a function of scaled position and momentum at $\sigma^2/x_\text{zp}^2 = \{5.5, 15, 30\}$ indicated by the three vertical dashed black lines and inside of each inset. Due to the scale invariance of the system, see Appendix~\ref{app:scale_invariance}, axis tick values in the inset are omitted.}
    \label{fig:robustness}
\end{figure}

To better understand how decoherence impacts the distinguishability between quantum and classical cubic phase states, in Fig.~\ref{fig:robustness} we examine three indicators: interference visibility, Wigner function negativity, and Jeffreys divergence. The latter, defined as $\jeffreysdiv{\pdfqt}{\pdfcl} \equiv \relentropy{\pdfqt}{\pdfcl} + \relentropy{\pdfcl}{\pdfqt}$, quantifies the statistical distance between the quantum and classical position distributions. For clarity, each measure is normalized to its value at $\sigma^2 = \zppos^2$.
At low decoherence values (e.g., $\sigma^2 \approx 5.5\zppos^2$), all three measures successfully distinguish between the classical and quantum hypotheses. This can be seen in the leftmost inset, which shows interference visibility, Wigner negativity, and a clear difference between the quantum and classical distributions. As decoherence increases, visibility drops to zero around $\sigma^2 \approx 13\zppos^2$. In contrast, Wigner function negativity persists to significantly higher values of $\sigma^2$, highlighting the presence of non-contextual quantum features~\cite{Delfosse2015}. Ultimately, Jeffreys divergence emerges as the most robust indicator, remaining nonzero up to $\sigma^2 \gtrsim 30\zppos^2$. Its resilience to decoherence arises because, as seen in Eq.\eqref{eq:low_pass}, decoherence acts as a low-pass filter, exponentially suppressing the high-frequency interference fringes that the visibility test relies on. In contrast, the likelihood ratio test leverages information from the entire distribution. Thus, distinguishing quantum from classical dynamics in the presence of decoherence is analogous to inferring the shape of an object concealed beneath a cloth. As decoherence increases, the cloth thickens, making the object harder to discern, but the underlying features remain and can therefore be tested.

\subsection{Polynomial vs. Exponential Scaling of $N_\star$}
\label{sec:universality}
Since the exponential advantage of the likelihood ratio test is based on the low-pass filtering of high-frequency features, the contrasting behavior observed in Fig.~\ref{fig:2} generalizes beyond cubic phase states under two conditions: (i) the probability distribution exhibits a separation of length scales, consisting of a slowly varying envelope and rapidly oscillating interference fringes; and (ii) noise and decoherence can be modeled as a Gaussian convolution with a kernel of width $\sigma^2$ (c.f. Eq.~\eqref{eq:low_pass}). We reiterate that states fulfilling (i) naturally arise in the WKB approximation and related semiclassical constructions \cite{Berry1972, LandauLifshitz1977, Heller2018}, whereas (ii) provides an accurate approximation of decoherence effects for systems evolving in wide non-harmonic potentials~\cite{RieraCampeny2024}.

The number of measurements $N_\star$ required to falsify classical mechanics at a significance level $\alpha = \Phi(-a)$ and statistical power $1-\beta = \Phi(b)$ is determined by the distance between the mean values of the test statistic relative to their variances (see App.~\ref{app:details_numerics} for details), that is
\begin{equation}\label{eq:N_star_main}
    N_\star = \left[\frac{b \,\sqrt{\std_1(Z)} + a \sqrt{\,\std_0(Z)}}{\langle Z \rangle_1 - \langle Z \rangle_0}\right]^2.
\end{equation}
Here, $\langle Z \rangle_s$ and $\std_s(Z) = N\times\text{Var}_s(Z)$ denote the expectation value and the $N$-independent part of the variance in the large $N$ limit under hypothesis $s \in \{0, 1\}$, respectively, and $\Phi(x)$ is the cumulative distribution function of the normal distribution.

Evaluating Eq.~\eqref{eq:N_star_main} for both test statistics $Z$ and the class of states defined by conditions (i) and (ii) reveals the contrasting \emph{exponential} versus \emph{polynomial} scaling. Here we sketch the main steps of the derivation in Appendix~\ref{app:universality}.

For the likelihood ratio test ($Z = \Lambda$), we assume the probability distribution $p_s(x)$ admits a \textit{Gram-Charlier A} series, valid for distributions that decay at least exponentially in $x^2$ as $|x|\to\infty$~\cite{Wallace1958}. This series expresses the distribution in terms of Hermite polynomials $H_n(x)$ with coefficients $c_s(n)$ defined in terms of Bell polynomials of the hypothesis-dependent cumulants. In the limit of large $\sigma^2$, one arrives at $N_\star \approx (a+b)^2\sigma^2/(\mu_1-\mu_0)^2$, when the hypotheses differ in the mean values $\mu_s$, while $N_\star \approx 2(a+b)^2\sigma^4/(\sigma_1^2-\sigma_0^2)^2$ for equal mean values but different variances $\sigma_s$. When the first two cumulants are hypothesis independent, a case often encountered in the systems of interest~\cite{RieraCampeny2024}, orthogonality of the Hermite polynomials in the expansion leads to
\begin{equation}\label{eq:poly_scaling}
    N_\star \approx (a+b)^2 \left[\sum_{n \ge 3} \frac{[c_1(n) - c_0(n)]^2}{n! \sigma^{2n}}\right]^{-1},
\end{equation}
giving polynomial scaling of leading order $(\sigma^{2})^d$ and $d = \min\lbrace n\geq 3\,|\,c_1(n)-c_0(n)\neq0\rbrace$. Conversely, for the visibility test ($Z = V$), condition (i) ensures that the interference fringes oscillate rapidly, with a length scale $1/k_0$, as compared to the width of the Gaussian noise kernel. Therefore, the convolution in condition (ii) suppresses the fringe contrast exponentially by a factor $\exp(-k_0^2 \sigma^2)$. Given that the visibility is defined in terms of bin counts that follow a multinomial distribution on bins of size $\Delta x$ one can show that
\begin{equation}\label{eq:exp_scaling}
    N_\star \approx \frac{\exp(k_0^2 \sigma^2)}{2v^2} \left[\frac{b}{\sqrt{A_1(x_0) \Delta x}} + \frac{a}{\sqrt{A_0(x_0) \Delta x}}\right]^2.
\end{equation}
Here, $v$ is the fringe visibility, and $A_s(x_0)$ is the value of the hypothesis-dependent envelope of $p_s(x)$ evaluated in the vicinity of the fringe position $x\approx x_0$. The comparison between the polynomial scaling of Eq.~\eqref{eq:poly_scaling} and the exponential scaling of Eq.~\eqref{eq:exp_scaling} formally demonstrates that the likelihood ratio test yields an exponential reduction in $N_\star$ relative to the visibility test, for all dynamical protocols satisfying conditions (i) and (ii).

\section{Conclusion}
\label{sec:conclusion}
In summary, we have introduced a likelihood ratio-based hypothesis testing framework that uses position measurements to certify quantum mechanics with finite data. Our analysis shows that this method can outperform the standard visibility test under realistic conditions relevant to levitated nanoparticle experiments. Specifically, for equivalent statistical significance and power, it distinguishes classical from quantum dynamics using exponentially fewer measurements and, moreover, can succeed even when decoherence fully suppresses quantum interference visibility. That is, we have shown that while finite visibility is sufficient and aesthetically pleasing for certifying quantum mechanics in interference experiments, it is, from a hypothesis testing perspective, not necessary. These results pave the way for certifying quantum superpositions of macroscopic particles under less restrictive experimental conditions.

\section*{Data Availability}
The code to reproduce the statistical analysis and figures is openly available~\cite{ZenodoRepo}.

\begin{acknowledgments}
We thank the Q-Xtreme synergy group for fruitful discussions. This research has been supported by the European Research Council (ERC) under the grant agreement No. [951234] (Q-Xtreme ERC-2020-SyG). ARC acknowledges funding from the European Union’s Horizon 2020 research and innovation programme under the Marie Skłodowska-Curie grant agreement No. [101103589] (DecoXtreme, HORIZON-MSCA-2022-PF-01-01).
\end{acknowledgments}

\section*{Author Contributions}
A.R.C. and P.M. developed the statistical certification framework and derived the analytical results. P.M. performed the numerical analysis. O.R.I. originated the idea and supervised the project. All authors discussed the results and contributed to the writing and editing of the manuscript.

\appendix 

\section{Characteristic Function of Cubic Phase States}\label{app:cubic_states}
In this Appendix, we derive the analytic expression for the characteristic function of the cubic phase states.

To obtain the characteristic function of a (position) cubic phase state, we evolve the initial Wigner function $\wigner_\mathrm{i}(\pos,\mom)$---that of an uncorrelated Gaussian state with variances $\variance_\pos$ and $\variance_\mom$---from just before $t_i = t_\mathrm{p}^{-}$ to just after $t_f = t_\mathrm{p}^{+}$ a cubic pulse. More precisely, the evolution is driven by a time-dependent potential of the form $\pot_{\mathrm{cubic}}(\pos,t) = -\hbar(\gamma/2) ( \pos/\zppos)^3 \delta(t - t_\mathrm{p})$, where $\gamma$ is a dimensionless parameter that sets the strength of the pulse. The dynamics follow Eq.~\eqref{eq:wigner_equation} with $s = 0$ or $s = 1$. Note that for the classical case, the $\hbar$ in $\pot_{\mathrm{cubic}}(\pos,t)$ is only a conventional choice, and does not imply that the dynamical laws are classical or quantum. The equation can be solved analytically for both the classical and the quantum hypothesis. One obtains the characteristic function
\begin{align}
    \charfunc_s(\kpos,\kmom) =& \exp\left\{i s \frac{\gamma}{3} \left(\zpmom \kmom \right)^3 - \frac{(\variance_\mom/\zpmom^2)}{2} \left(\zpmom \kmom \right)^2 \right. \nonumber\\
    &\left.-(\variance_\pos/\zppos^2)\frac{(\zppos \kpos)^2}{2}\frac{1}{1-2i\gamma (\variance_\pos/\zppos^2) \zpmom \kmom} \right\}\nonumber\\
    &\times 1/\sqrt{1-2i\gamma (\variance_\pos/\zppos^2) \zpmom \kmom},
\label{eq:charfunc_cubic_cl}
\end{align}
corresponding to the Fourier transform of $\wigner_s(\pos,\mom,t_f)$. Setting $\kpos = 0$ and $\kmom\mapsto k$, one obtains the characteristic function in the form displayed in Eq.~\eqref{eq:charfunc_cubic_s}. Note that under Gaussian dynamics, the shape of the state will be preserved while the position and momentum variables can be mixed by symplectic transformations~\cite{Serafini2017}.

\section{Scale Invariance}\label{app:scale_invariance}
In this Appendix, we discuss the scale invariance properties of the position probability distributions for cubic phase states.

The probability distribution associated to Eq.~\eqref{eq:charfunc_cubic_s} in the main text, exhibits the scale invariance $\pdf_s(\signal;\param_1,\param_2,\param_3) = |\lambda| \pdf_s(\lambda \signal;\lambda\param_1,\lambda^2\param_2,\lambda^3 \param_3)$, for $\lambda \neq 0$. Since the random measurement noise $\variance_R$ is unaffected by this transformation, only the parameters $\theta_k$ scale, the relative impact of measurement noise can be reduced by increasing $\lambda$. Specifically, the effective variance of the Gaussian convolution above is transformed as $\sigma^2 \mapsto \lambda^2 (\theta_2 - \param_3/\param_1) + \variance_R,$ showing that the contribution of $\variance_R$ becomes negligible for large $\lambda$. Physically, such scaling corresponds to a squeezing transformation, which can be implemented by evolving the particle under an inverted harmonic potential~\cite{Romero-Isart2017, Neumeier2024}.
This property is used to produce the Figures displayed in the main text. 

\section{Definition of the Visibility Test Statistic}\label{app:visibility_test}
In this Appendix, we provide a rigorous definition of the interference visibility statistic used in our analysis, including the procedure for determining the integration bins corresponding to the maximum and minimum of the interference pattern.

In Eq.~\eqref{eq:visibility}, we have introduced the definition of the visibility $\vis$ as a function of the counts $\nmax$ and $\nmin$. To be precise, we define it as follows. Let $\posmax$ and $\posmin$ be the values of the second maximum and first minimum of the quantum position distribution $\pdfqt(\datapoint)$, which are at a distance $\Delta \equiv |\posmax - \posmin|$, and which we assume exist. Then, we define two intervals $\imax = [\posmax -\Delta/2, \posmax + \Delta/2]$ and $\imin = (\posmin -\Delta/2, \posmin+ \Delta/2]$, centered around $\posmax$ and $\posmin$. Given the dataset $\{\datapoint_1, \cdots, \datapoint_\nm\}$, we define $\nmax = \sum_{i=1}^\nm \indicator(Y_i\in\imax)$ and $\nmin = \sum_{i=1}^\nm \indicator(Y_i\in \imin)$ as the number of counts on the intervals $\imax$ and $\imin$, respectively. Here, $\indicator(c)$ is the indicator function that gives one if the condition $c$ is true and zero otherwise. The number of counts in the $\imax$ and $\imin$ bins are part of a multinomial distribution, whose statistical properties---for instance, the mean values, variances, and covariances---can be evaluated.
To conclude, we note that in the definition of Eq.~\eqref{eq:visibility}, we omit the conventional absolute value, since the classical distribution could have a negative visibility. In such cases, including the absolute value would artificially degrade the performance of the test. Also, we remark that for large decoherence it is not guaranteed that $\posmax$ and $\posmin$ exist. In that case, it is customary to set the visibility to zero. This makes the visibility of the quantum distribution decay continuously to zero. 

\section{Physical Parameters of the Proposal}\label{app:physical_parameters}
In this Appendix we show how the physical parameters discussed in the proposal of Ref.~\cite{Neumeier2024} relate to the parameters of the characteristic function in Eq.~\eqref{eq:charfunc_cubic_s}. 

The proposal consists of a five-step protocol implemented with a single charged dielectric silica nanoparticle of mass $m$. Each step $0 \leq i \leq 4 $ is fully characterized by its duration $t_i$, decoherence rate $\Gamma_i$, and potential $\pot_i(\pos) = -\mass\trapfreq_i^2/(2k^2)[\cos^2(k \pos - \phi_i) - k \pos \sin(2\phi_i)]$, with frequency $\trapfreq_i$ and phase $\phi_i$. The potential can be realized by optical standing wave pulses with wavenumber $k$ and a linear electric potential. At step 0 the particle is initialized by cooling it to a thermal state of mean phonon occupation $\nbar$ in a harmonic potential with frequency $\trapfreq_0 = \trapfreq$. After that, the protocol follows a sequence of free evolutions and evolutions in cubic and inverted potentials with fine tuned parameters to generate a cubic phase state as in Eq.~\eqref{eq:charfunc_cubic_s}.
Following the derivation of Ref.~\cite{Neumeier2024} it can then be shown that the relation between the physical parameters and the parameters of the characteristic function are given by Tab.~\ref{tab:sm_params}, where
\begin{align}\notag
    \label{eq:nu_sm}
    \frac{\tilde{a}}{\trapfreq^2} &= 4\Gamma_2 t_2t_1^2 + \frac{4\Gamma_1t_1^3}{3}+\frac{\exp(2\Omega_4 t_4)(4\Gamma_3 t_3^3/3 + 2\Gamma_4/\Omega_4^3)}{4\lambda^2}.
\end{align}

\begin{table}[b]
\caption{Relation between the physical parameters discussed in the proposal of Ref.~\cite{Neumeier2024} and the parameters of the characteristic function in Eq.~\eqref{eq:charfunc_cubic_s}. Numerical values are obtained by inserting the values in Table I of Ref.~\cite{Neumeier2024}, including the assumption of negligible measurement noise $\variance_R = 0$.}
\begin{ruledtabular}
\begin{tabular}{c|c|c}
  Parameter & Physical Parameters &Value \\
  \hline
  $\param_1/(\lambda\zppos)$ & $ (2\nbar + 1) (k \zppos) (\trapfreq t_1)^3  $ & $69.04$ \\
  $\param_2/(\lambda\zppos)^2$ & $ (2\nbar + 1 + \tilde a)$ &  $6.001$\\
  $\param_3/(\lambda\zppos)^3$ & $ (k \zppos)(\trapfreq t_1)^3$ &  $34.52$\\
  $\lambda$ & $-\cosh(\Omega_4 t_4)(t_3/t_1)-\sinh(\Omega_4 t_4)/(\Omega_4 t_1)$ & $-59.67$
\end{tabular}
\end{ruledtabular}
\label{tab:sm_params}
\end{table}

\section{Details on Numerical Evaluation}\label{app:details_numerics}
In this Appendix we show how we determine the statistical power $1-\beta$ and number of measurements $\nmt$ required to falsify classical mechanics for $M$ runs per hypothesis. 

First we evaluate the mean value $\mu$ and standard deviation $\sqrt{\mathrm{Var}_0(Z)}$ of the test statistic $Z$ for the null hypothesis. The threshold is then set to $Z_\star = \mu + 5\sqrt{\mathrm{Var}_0(Z)}$ which corresponds to a significance level $\sig \approx 2.86 \times 10^{-7}$. This allows us to evaluate $M_>\equiv \sum_{i=1}^M \indicator(Z_i>Z_\star)$, that is the total number of runs with a test statistic $Z_i$ that lies above the threshold.

Using $M_>$ we evaluate the Wilson score interval, which guarantees that the statistical power $1-\beta$ is with a probability $1-\epsilon$ within the interval $[w_-(\epsilon),w_+(\epsilon)]$, where
\begin{align}
    w_\pm(\epsilon) &= \frac{M_>+ z_\epsilon^2/2}{M+z_\epsilon^2}\pm \frac{z_\epsilon/2}{M+z_\epsilon^2} \sqrt{\frac{4  (M-M_>)M_>}{M} + z_\epsilon^2},
\end{align}
with $z_\epsilon = \Phi^{-1}(1-\epsilon/2)$, where $\Phi(x)=\int_{-\infty}^x \exp(-s^2/2)/\sqrt{2\pi} \,ds$ denotes the cumulative distribution function of the normal distribution. For a conservative estimate we evaluate the power at the lower interval and $\epsilon=0.05$, i.e. $1-\beta = w_-(0.05)$. Note that $w_-(\epsilon)\leq 1/(1+z_\epsilon^2/M)$, which requires a minimum number of measurements to ensure $1-\beta \geq 99.73\%$ which is why in Fig.~\ref{fig:2}b the empirical $\nmt$ does not drop below 1000.

In order to determine $\nmt$ we consider a $\pm 5\%$ error margin for the parameters. Specifically, we fix the parameters $\theta_i$ used to evaluate the likelihood ratio but sample the data from a probability distribution with parameters within the interval $[0.95 \theta_i, 1.05 \theta_i]$. This accounts for potential uncertainties and ensures robustness in the estimation of $\nmt$. 

Finally, the shaded regions can be obtained by first evaluating the mean value $\langle Z \rangle_s$ and variance $\text{Var}_s(Z)$ under hypothesis $s\in \lbrace 0,1\rbrace$ in the large $N$ limit. We reject the null hypothesis when $Z_i > \langle Z \rangle_0 + a \sqrt{\text{Var}_0(Z)}$ which for Gaussian distributed test statistics corresponds to a significance level $\alpha = \Phi(-a) $. The statistical power reads
\begin{align}
    1-\beta &=1 - \Phi\left(\frac{\langle Z \rangle_0 + a \sqrt{\text{Var}_0(Z)} - \langle Z \rangle_1}{\sqrt{\text{Var}_1(Z)}}\right),
\end{align}
and exceeds a value $1-\beta\geq 1-\Phi(-b)= \Phi(b)$ if $\langle Z \rangle_1-\langle Z \rangle_0 \geq a \sqrt{\text{Var}_0(Z)}+b\sqrt{\text{Var}_1(Z)}.$ Using $\text{Var}_s(Z) = \std_s(Z)/N$ allows us to derive the number of measurements $\nmt$ required to falsify classical mechanics in the large $N$ limit, namely
\begin{equation}\label{eq:N}
    N_\star = \left[\frac{b \,\sqrt{\std_1(Z)}+a \sqrt{\,\std_0(Z)}}{\langle Z \rangle_1-\langle Z \rangle_0}\right]^2.
\end{equation}

\section{Derivation of Scaling Laws}\label{app:universality}

Here we demonstrate that the contrasting \emph{exponential} versus \emph{polynomial} scaling of the visibility and likelihood ratio test with noise generalizes beyond cubic phase states under the following conditions: (i) the probability distribution exhibits a separation of length scales, consisting of a slowly varying envelope and rapidly oscillating interference fringes; and (ii) noise and decoherence can be modeled as a Gaussian convolution. 

To be precise, condition (i) implies that we can write down the probability distribution of the two hypothesis as 
\begin{equation}\label{eq:toymodel}
 p_s(x)=A_s(x)\left[1+s v\cos\left(k_0x\right)\right],
\end{equation}
where $A_s(x)$ is the slowly varying envelope and $s$ distinguishes the quantum ($s=1$) from the classical ($s=0$) hypothesis. The quantum distribution contains interference fringes with visibility $v$ and spatial frequency $k_0$. On the other hand, condition (ii) implies that after noise and decoherence the probability distribution in position is given by the Gaussian convolution
\begin{equation}\label{eq:noise}
   (p_s*g_\sigma)(x)=\frac{1}{\sqrt{2\pi \sigma^2}}\int_{-\infty}^\infty \exp\left[-\frac{(x-x')^2}{2\sigma^2}\right]p_s(x')\,\mathrm{d}x',
\end{equation}
where $\sigma$ quantifies the noise strength. 

Our goal is to quantify how the number of measurements $\nmt$ required to certify quantum mechanics scales with the decoherence strength $\sigma$, for both the visibility $Z = \vis$ and the likelihood ratio $Z = \lrt$. To this end, we use Eq.~\eqref{eq:N} and compute the expectation $\langle Z \rangle_s$ and standard deviation $\std_s(Z)$ of the test statistic $Z$ under the two hypotheses $s=0$ and $s=1$.

We start with the likelihood ratio $Z = \lrt$. First, we assume that the distribution $p_s(x)$ admits a \textit{Gram-Charlier A} series, valid for distributions that decay at least exponentially in $x^2$ as $|x|\to\infty$~\cite{Wallace1958}, namely
\begin{equation}
   p_s(x)=\frac{1}{\sqrt{2\pi\sigma_s^2}}\exp\left[-\frac{(x-\mu_s)^2}{2\sigma_s^2}\right]\sum_{n\geq0}\frac{c_s(n)}{n!\sigma_s^n} H_n\left(\frac{x-\mu_s}{\sigma_s}\right),
\end{equation}
where $c_s(n)\equiv B_n(0,0,\kappa_{s}(3),\dots,\kappa_{s}(n))$ denote the complete Bell polynomials, $\kappa_s(n)$ is the $n$th cumulant of the distribution $p_s(x)$, and $H_n(x)$ denote the Hermite polynomials. Due to condition (ii), noise modifies the variance as $\sigma_s^2 \to (\sigma_s^2+\sigma^2$). To obtain the scaling, we evaluate $\langle Z \rangle_s$ and $\std_s(Z)$ by inserting the Gram-Charlier A series, expanding $\log(1+x)\simeq x$ and approximating the denominator of the resulting expression by a Gaussian, valid for large enough $\sigma$. When the hypotheses differ in the mean values, $N_\star \simeq (a+b)^2\sigma^2/(\mu_1-\mu_0)^2$, while for equal mean values but different variances $N_\star \simeq 2(a+b)^2\sigma^4/(\sigma_1^2-\sigma_0^2)^2$. When the first two cumulants are identical, orthogonality of Hermite polynomials leads to
\begin{equation}\label{eq:poly}
    N_\star  \simeq (a+b)^2 \left[\sum_{n\geq 3} \frac{[c_1(n)-c_0(n)]^2}{n!\sigma^{2n}}\right]^{-1},
\end{equation}
giving polynomial scaling with $(\sigma^2)^d$ and $d = \min\lbrace n\geq 3\,|\,c_1(n)-c_0(n)\neq0\rbrace$.

For the visibility test we return to Eq.~\eqref{eq:toymodel}. Condition (i) implies that $A_s(x)$ is slowly varying on the scale of $1/k_0$ and $\sigma$, and therefore $(p_s * g_\sigma)(x)\simeq A_s(x)\left[1+sv\exp(-k_0^2\sigma^2/2)\cos(k_0x)\right],$ showing exponential suppression of fringe contrast $v$. The visibility is defined in terms of bin counts that follow a multinomial distribution on bins of size $\Delta x$. Applying the delta method~\cite{Casella2002} we evaluate $\langle Z \rangle_s$ and $\std_s(Z)$ near the fringe position $x\simeq x_0$, obtaining
\begin{equation}\label{eq:exp}
    N_\star  \simeq \frac{\exp(k_0^2\sigma^2)}{2v^2} \left[\frac{b}{\sqrt{A_1(x_0) \Delta x}}+\frac{a}{\sqrt{A_0(x_0) \Delta x}}\right]^2,
\end{equation}
Finally, comparing Eq.~\eqref{eq:poly} and Eq.~\eqref{eq:exp} demonstrates that the likelihood ratio test yields an exponential reduction in $N_\star$ relative to the visibility test, for all dynamical protocols satisfying conditions (i) and (ii).

\bibliography{bibliography.bib}

\end{document}